\documentclass[aps,pra,twocolumn,superscriptaddress,amssymb,showpacs]{revtex4}
\usepackage{verbatim}
\usepackage{graphicx}

\begin{document}

\title{Chirped quasi-phase-matching with Gauss sums for production of biphotons}

\author{D.~A.~Antonosyan}
\email[]{antonosyand@ysu.am}

\affiliation{Yerevan State University, Alex Manoogian  1, 0025,
Yerevan, Armenia}\affiliation{Institute for Physical Researches,
National Academy of Sciences,\\Ashtarak-2, 0203, Ashtarak,
Armenia}

\author{A.~R.~Tamazyan}
\email[]{a.tamazyan@ysu.am}

\affiliation{Yerevan State University, Alex Manoogian  1, 0025,
Yerevan, Armenia}\affiliation{Institute for Physical Researches,
National Academy of Sciences,\\Ashtarak-2, 0203, Ashtarak,
Armenia}

\author{G.~Yu.~Kryuchkyan}
\email[]{kryuchkyan@ysu.am}

\affiliation{Yerevan State University, Alex Manoogian  1, 0025,
Yerevan, Armenia}\affiliation{Institute for Physical Researches,
National Academy of Sciences,\\Ashtarak-2, 0203, Ashtarak,
Armenia}

\begin{abstract}
We study the theory of linearly chirped  biphoton wave-packets produced in two
basic quasi-phase -matching configurations: chirped photonic-like  crystals and
aperiodically poled crystals. The novelty is
that these structures are considered as  definite assembles of nonlinear layers
that leads  to detailed description
of  spontaneous parametric down-conversion (SPDC) processes through the discrete
Gauss sums. We demonstrate that  biphoton spectra for chirped photonic crystals
involving a small number of layers consist from definite well-resolved spectral
lines. We also discuss  the forming of broadband spectra of signal (idler) waves
in SPDC for both configurations as number of layers increases as well as in
dependence of chirping parameters .

\end{abstract}


\maketitle

\section{INTRODUCTION}

One of the challenges in the fields of
quantum optics and quantum communications is generation of new sources of nonclassical  light, such as biphotons
 with controllable spectral and
temporal properties. The standard method for the generation of biphotons is
spontaneous parametric down-conversion (SPDC) \cite{HsOsh,MKl,Shih}, where
photon pairs are generated under action of a strong pump field interacting with a
 nonlinear crystal. In particular, of considerable interest is preparation of
 biphoton wave-packets with small correlation times between two-photons as well
 as with broad frequency spectra.  Correlated photon pairs with broad spectra
 are mainly generated by SPDC using method of quasi-phase-matching (QPM) that
 provides a feasible alternative to conventional phase matching for many optical
 parametric process applications. Several methods have been suggested and
 experimentally implemented in this direction for generation of biphotons,
 including SPDC in aperiodically poled crystals \cite{Cro,Crrs,nast,DnUr,Harris,Perina}.
  Studies in this direction allow us to understand three-photon interactions at
  scale of the shortest possible correlation time that is a single optical
  cycle \cite{Harris}. Thus, these investigations are interesting  from the fundamental point of view as well as  are useful for
  further applications in quantum technologies. QPM leading to photon pair
  production is effectively realized in layered nonlinear structures,
  particularly, in periodically-polled second-order crystals or
  photonic
  crystals. The applications have been done  for synthesis of twin photon
 states by manipulating overall group delay mismatches between interacting waves
 in a multilayered structures by using compensating dispersive
 effects \cite{Kly,SergTch,DiGi,Uren}, for production of entangled three-photon
 states in cascaded parametric processes \cite{WSAK,DTG,DOptCom}.

 As it has been
 recently experimentally demonstrated biphotons with broad spectra and ultrashort
 correlation time at a high emission rate can be generated using chirped QPM
 nonlinear crystals that involve nonlinear grating in the spatial coordinate
 along the direction of pump propagation with a nonuniform poling
 period \cite{NsCrs,kitaeva,SenHar,Imeshev}. In this case, phase-matching is
 achieved along the crystal in such a way that the phase-matching conditions
 vary longitudinally by use of QPM which leads to broadband biphoton.

 The theory of linearly chirped
   biphoton wave-packets is usually based on simple model allowing analytical
    description of linearly chirped periodically-poled crystals or optical fibers
     (see, \cite{NsCrs,kitaeva,SenHar,Imeshev}). In this model a phase-matching function of three-wave interaction $exp[i\Delta k(z)z] $ with a QPM grating with  linearly varying spatial frequency $\Delta k(z)= \Delta k- \alpha z$  has been considered. The interaction leads to  biphoton
spectral density in form of the  erfi(x) error function.
       Such phenomenological approach
       qualitatively explains the basis properties of chirped
       quasi-phase-matching, in this number: extremely wide spectral bandwidths
       of biphoton wave packets and the sharp temporal separation between the
       signal and idler photons comprising a pair. However, this approach has been unable
       to describe some  important details of chirped three-wave interactions.

In this paper, we present more detailed description of chirped structures for
generation of two-photon light via SPDC
from a cw pump in another way: as the definite assembly of nonlinear segments.
This approach provides more flexibility for designing of QPM gratings
 and allows us engineer the controllable phase relation between
      the various spectral components. On the whole it becomes possible to control  the frequencies and bandwidths
       of the signal and idler photons  by varying also the number of layers equally with the local poling period
       along the length of the crystal.  In this approach the total amplitude of
two-photon generation is calculated as the superposition of partial amplitudes
corresponding to each layer with local spatial chirp frequency. As frequencies
of signal and idler photons are varied with different layers a much broader range
of photon frequencies is formed due to the superposition. Such detailed analysis
of chirped QPM allows considering interference and transition effects in SPDC
stipulated by definite numbers of layers. In this approach the
quasi- phase-matching function of SPDC is expressed through so called Gauss
sums instead of the continuous error function that is appeared in the
phenomenological approach. It should be mentioned that during recent years it has been seen an
impressive number of experiments implementing Gauss sums
\begin{equation}
S_{N}(\zeta)=\sum_{m=-M}^{M}W_{m}e^{2\pi i(m+\frac{m^{2}}{N})\zeta}
\end{equation}
 in physical systems, particularly,  for number factorization schemes. These systems range from nuclear magnetic resonance (NMR)
methods \cite{MMASch,RajPeng,Suter} via cold atoms \cite{GWSc} and Bose-Einstein
condensates (BECs) \cite{Kumar}, tailored ultrashort laser
pulses \cite{SchGir,Weber} to classical light in a multi-path Michelson
interferometer \cite{Tamma,TammaSch} (for an introduction to this field,
see \cite{Wolk,Merkel}). In addition to these results here we present new
quantum systems for implementation of Gauss sums to factor numbers.

We investigate two structures: chirped layered photonic crystals
and aperiodically poled layered crystals. In the phenomenological
approach in which the  phase-matching function is described by the
continuous error function both structures give the same results
distincting only on the physical  parameters. However, detailed
description of these chirped structures  shows  important
differences between them.

The paper
 is organized as follows: In Sec. II, the brief description of SPDC in multilayered media is presented. Sec. III is devoted to chirp  in layered photonic-like crystals. The production of broadband biphotons in aperiodically poled layered
 crystals is described in Sec. IV.  We summarize our results in Sec. V.

\section{SPDC in Multilayered Media: Brief Description}

In this section we briefly describe generation of two-photon light
via SPDC in
one-dimensional $\chi^{(2)}$ media consisting of layers with
different coefficients of nonlinearity and refractive indices \cite{Kly}, \cite{Uren}, \cite{WSAK}.  We
consider collinear, type-II  QPM configurations for generation of photons at the same frequency.
Since the generated photons are collinear, their directions cannot be used to distinguish the two photons.
Thus, we assume that photons in pairs have different polarizations, but omit the polarization indexes below.
 Three-wave interaction Hamiltonian is expressed as the sum of
 interactions in each layer in terms of the electric fields for the
 m-th layer, $E_{0m}(z,t), E_{jm}^{-}(z,t), E_{jm}^{+}(z,t)$, j=(s,i),
 \begin{equation}
 H(t)=\sum_{m}{\int_{z_{m}}^{z_{m+1}}{dz\chi^{(2)}(z)E_{0m}^{*}(z,t)E_{im^{(-)}}E_{sm}^{(-)}(z,t)+h.c.}}
 \end{equation}
 Here  $\chi
 ^{(2)}(z)$ is the second-order susceptibility,
 $E_{0m}(z,t)$ represents classical laser field at the frequency $\omega_{0}$ and $E_{im}^{(-)}(z,t)$,
 $E_{sm}^{(-)}(z,t)$ represent the positive-frequency parts of the
 fields of the subharmonics centered at the frequencies  $\omega_{i}=\frac{\omega_{0}}{2}+\Omega, \omega_{s}=\frac{\omega_{0}}{2}-\Omega$.

The two-photon state can be written as
\begin{equation}
|\psi\rangle=\int \int d\omega_{s}d\omega_{i}
\Phi(\omega_{s},\omega_{i})a^{+}(\omega_{s})a^{+}(\omega_{i})|0\rangle,\label{2phs}
\end{equation}
where $a^{+}(\omega_{s})$ and $a^{+}(\omega_{i})$ are the creation
photon operators for modes with frequencies $\omega_{s}$ and
$\omega_{i}$, $\omega_{0}=\omega_{s}+\omega_{i}$, $|0\rangle$ is a
vacuum state of the signal and the idler fields and $\Phi(\omega_{s},\omega_{i})$ is the spectral
amplitude of two photon radiation. Two-photon amplitude for multilayered structures has been presented
in many papers \cite{Kly}, \cite{Uren}, (see, also \cite{WSAK, DTG, DOptCom}). It has been
calculated as the sum of definite partial amplitudes in the
general form given by the product of the pump envelope function
$E_{0}(\omega_{0})$ and the phase matching function $F(\Delta k)$
\begin{equation}
\Phi(\omega_{s},\omega_{i})=\frac{-2\pi
i}{\hbar}E_{0}(\omega_{s}+\omega_{i})F(\Delta k),\label{poqrEq}
\end{equation}
\begin{eqnarray}
F=\sum_{m}{l_{m}\chi_{m}F_{m}},~~~~~~~~~~~\nonumber\\
F_{m}=e^{-i\left(\varphi_{m}+\frac{\Delta
k_{m}l_{m}}{2}\right)}sinc\left(\frac{\Delta
k_{m}l_{m}}{2}\right)\nonumber\\
\varphi_{m}=\sum_{n}^{m-1}l_{n}\Delta k_{n},~~~~~~ \varphi_{1}=0.
\label{KlyshkoEq}
\end{eqnarray}
Here, $\chi_{m}$ is the second-order susceptibility in the $m$-th
layer, $l_{m}=z_{m+1}-z_{m}$ is the length and $\Delta
k_{m}=k^{m}_{0}-k^{m}_{s}-k^{m}_{i}$ is the phase mismatch vector
for the $m$-th layer,
$k^{m}_{j}(z,\omega_{j})=\frac{\omega_{j}}{c}n_{m}(z,\omega_{j})$,
$n_{m}(\omega)$ is the corresponding refractive index of the
medium at the given frequency that describes the effects of dispersion on the properties of photon pairs.  In this case the probability of twin-photon
generation is calculated as
\begin{eqnarray}
|\Phi(\omega_{s},
\omega_{i})|^{2}=\frac{(2\pi)^{2}}{\hbar^{2}}|E_{0}|^{2}\times|F|^2,\nonumber
\\|F|^2=\sum_{m=1}^{N}{|F_{m}|^{2}}+
2Re\big(\sum_{m_{1}<m_{2}}{F_{m_{1}}F_{m_{2}}^{*}}\big)\label{PhiSqr}
\end{eqnarray}
in the form which clearly demonstrates that the interfering
probability amplitudes $A_{m}\sim E_{0}F_{m}$ of two-photon
generation in each layer contribute to the total probability.
Further, for convenience we expand the wave vectors in Taylor series around the exact quasi-phase-matching frequency and take into  consideration only the zero- and first-order terms: $k_{s}=k_{s}\left(\frac{\omega_{0}}{2}\right)-\Omega k'_{s}$, $k_{i}=k_{i}\left(\frac{\omega_{0}}{2}\right)+\Omega k'_{i}$,
where $k'_{i,s}$ are the first derivatives of the dispersion law
evaluated at $\frac{\omega_{0}}{2}$, related to the group
velocity.  This approximation means that the spectrum of
biphotons is not too broad compared to the difference of group
velocities of the subharmonics.

\section{Chirp in photonic-like crystals}

Recently, non linear photonic crystals have been used in the
context of the process of spontaneous parametric down-conversion.
In particular, the  semiconductor-based nonlinear one-dimensional
photonic crystals have been explored for phase-matching in
generation of polarization-entangled photon pairs \cite{bouw1},
\cite{bouw2}. It  has been demonstrated  that one-dimensional
photonic crystal are very effective for generation of photon pair
due to field localization in such structures \cite{pc1},
\cite{pc2}, \cite{pc3}. The production of photon pairs with
engineered spectral entanglement properties in one-dimensional
nonlinear photonic crystals has been shown in  \cite{pc4}.

In this section, we study  one-dimensional nonlinear
photonic crystals for chirp configuration.
In the  scheme proposed generation of photon pairs by SPDC in
$\chi^{\left(2\right)}$ layered structure characterized by a chirp
in its linear optical properties. Considering the general equation
(\ref{KlyshkoEq}), we assume that layers have the equal lengths
$l_{m}=l$ as well as the equal susceptibilities
$\chi_{m}=\chi_{0}$, while refractive indexes of the pump
$n(\omega_{0},z)=n_{0}(z)$ and the generated waves
$n(\omega_{j},z)=n_{j}(z)$ ($j=s,i$) have linear dependence on
the coordinate and are given by

\begin{eqnarray}
n_{0}(z)=\left\{ \begin{array}{ll}
   n_{0}, & 0<z<l;\\
   n_{0}-ml\beta_{0} & ml<z<(m+1)l,\end{array}\right.\ \nonumber\\
n_{j}(z)=\left\{ \begin{array}{ll}
   n_{j}, & 0<z<l;\\
   n_{j}-ml\beta_{j} & ml<z<(m+1)l,\end{array}\right.\ \label{refind}
\end{eqnarray}
where $\beta_{0}$ and $\beta_{j}$ ($j=s,i$) are chirp parameters
for the refractive indexes of the pump and subharmonic waves,
correspondingly. In this case, the phase-mismatch vector for the
$m$-th layer has a linear chirp of the following form $\Delta k_{m}=\Delta k-\alpha( m-1)l$,
where $\Delta
k=k_{0}(\omega_{0})-k_{s}(\omega_{s})-k_{i}(\omega_{i})$
is the phase mismatch vector at the first layer, and
$\alpha=\frac{\omega_{0}}{c}(\beta_{0}-\frac{\beta_{s}}{2}-\frac{\beta_{i}}{2})$
is the spatial chirp parameter.

Taking into account (\ref{refind}) we present the general
expression (\ref{KlyshkoEq}) as

\begin{eqnarray}
F(\Delta k)=l\chi_{0}e^{i\frac{l\Delta
k}{2}}\sum^{N}_{m=1}{F_{m}},~~~~~~~~~~~~~ \nonumber
\\F_{m}=e^{-i(ml\Delta
k-\frac{\alpha (m-1)^{2}l^{2}}{2})}\times sinc\left(\frac{(\Delta
k-\alpha( m-1)l)l}{2}\right),
\end{eqnarray}
where $sinc(x)=\frac{sin(x)}{x}$.

It is easy to see that in the frame of this presentation, the phase
matching function is given by the Gauss sum, in spite of the error
functions that are appeared in the phenomenological approach \cite{Harris,kitaeva}.
Below, we calculate these sums considering  the phase-mismatch function  as $\Delta
k=\Delta k_{0}+\Omega D$, where $\Delta k_{0}=k_{0}-k_{s}(\frac{\omega_{0}}{2})-k_{i}(\frac{\omega_{0}}{2})$ and $D=k'_{s}-k'_{i}$  is temporal walkoff between signal and
idler modes,  assuming that $\Delta k_{0}$ satisfies the QPM condition.
 In order to illustrate the broadening of the spectrum and the other features of
obtained phase-matching function  we
have calculated the squared amplitude of the phase-matching
function $|F(\Omega)|^{2}$  on the base of the formula
(\ref{PhiSqr}) and the wave vector dispersion expansion as

\begin{widetext}
\begin{eqnarray}
|F(\Omega)|^{2}=l^2\chi^{2}_{0}\Bigg[\sum^{N}_{m=1}{sinc^{2}\left(\frac{(D\Omega+\Delta k_{0}-\alpha
(m-1)l)l}{2}\right)}+ 2\sum^{N-1}_{m=1}\sum^{N-m}_{p=1}\Bigg(
sinc\left(\frac{(D\Omega+\Delta k_{0}-\alpha (m-1)l)l}{2}\right)\times\\
\nonumber \times sinc\left(\frac{(D\Omega+\Delta k_{0}-\alpha
(m+p-1)l)l}{2}\right)\Bigg)\cos\left(p(D\Omega+\Delta k_{0})l-\alpha
p\bigg(m-1-\frac{p}{2}\bigg)l^{2}\right)\Bigg].\label{PMFSq}
\end{eqnarray}
\end{widetext}

\begin{figure}
\includegraphics[width=9cm]{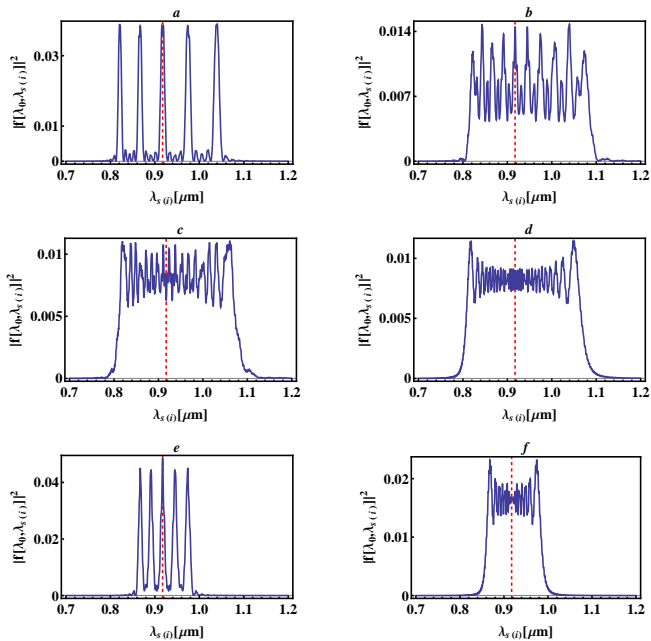}
\caption{Biphoton spectral density in dependence
on the wavelengths of signal (idler) field for the crystals with the
length $L=8000 \mu m$, dispersion parameter $B=0.3$, chirp
parameter $\alpha=1.2 \times 10^{-5}\mu m^{-2}$ and different numbers of
layers: $N=5$, $\Delta k_{0}=3\alpha l$ (a); $N=10$, $\Delta
k_{0}=5\alpha l$ (b); $N=20$, $\Delta k_{0}=10\alpha l$ (c) and
$N=80$ ,$\Delta k_{0}=40\alpha l$ (d). The cases of the chirping parameter
$\alpha=6\times 10^{-6} \mu m^{-2}$:   $N=5$, $\Delta k_{0}=3\alpha
l$  (e); $N=80$, $\Delta
k_{0}=40\alpha l$,  (f). The red dashed
lines correspond to the wavelength $2\lambda_{0}=0.916 \mu m$.}
 \label{Sp4}
\end{figure}

We present a detailed analysis of normalized  biphoton spectral
density $|f|^{2}=\frac{|F|^{2}}{L^{2}\chi_{0}^{2}}$ in dependence
on the wavelength of signal (idler) mode. In this way, we select
in the  Fig. (\ref{Sp4})  the results for typical values of pump
wave length $\lambda_{0}=0.458\mu m$, the chirp parameter
$\alpha=1200 cm^{-2}$, $B=cD=0.3$ and crystal with the total
length $L=0.8 cm$ taken from the experimental results on
LiTaO$_{3}$ \cite{kitaeva}. Note, that our approach allows to
consider the general case of arbitrary number of layers on the
base of the formula (9). To develop comparatively analysis  we
represent the squared amplitude of the phase matching function for
the crystals with the fixed  total length $L$, but with the
various number of domains, particularly,
 for the cases of small and large numbers of  layers.

The  results are depicted in
 Figs.\ref{Sp4}(a,b) for $N=5$ and $N=10$ layers and the phase-matching conditions  $\Delta k_{0}=3\alpha l$
and $\Delta k_{0}=5\alpha l$, correspondingly. As we see, in these
cases the biphoton spectra consist from definite separated
spectral peaks corresponding to the number of domains.  The
frequencies of the spectral lines are given by
 $\Omega_{m}=[\alpha (m-1)l-\Delta k_{0}]D^{-1}$ according to the formula (9) provided that
the spectral  lines are resolved i.e. their widths are less
than the intervals between them in the range of the wave-length
broadening.  Thus, there are  critical ranges of domains number
when the spectral lines are not resolved and chirp leads to the
broadening of the biphoton  spectra.

If domains' number is increasing, the chirp is displayed as a
broadening of the  biphoton spectra as it is depicted in
Figs.\ref{Sp4}(c,d) for $N=20$ and $N=80$ layers and the
phase-matching conditions  $\Delta k_{0}=10 \alpha l$ and $\Delta
k_{0}=40\alpha l$, respectively. One can see that the spectrum of
the biphoton field is quite broad (from 900 nm to 1300 nm wavelength )
 and has nearly rectangular shape already for $N=80$ (see,
Fig.~\ref{Sp4}(d)). For further increasing of the number of
domains the interference picture become smooth and the results
qualitatively coincidence with the analogous results obtained on
the base of simple model \cite{NsCrs, kitaeva}.

It should be noted
that the length of biphoton spectral shape does not depend from
the number of layers for fixed total length of nonlinear media but
is only determined by the chirping parameter. We illustrate this statement
on Fig.\ref{Sp4} (e,f) by consideration of the other chirping
parameter $\alpha=0.000006 \mu m^{-2}$ that is two times less than the previous one, for N=5 and N=80
layers, correspondingly. As we see, increasing of the chirping parameter leads to the increase of the intervals between the spectral peaks for the case of small numbers of layers. For the case of large layers' numbers this increasing leads to the increase of the range of the  wave-length spectral broadening.

\section{Chirp in aperiodically poled crystals}

Recently,  a aperiodic poling have been used for various applications. It has been provided not only for compensation of a natural phase mismatch but has been allowed to tailor the properties of emitted photon pairs using nonlinear domains with variable lengths (chirped periodical poling). Domains of different lengths in an ordered structure allow an efficient nonlinear interaction in an ultra-wide spectral region extending typically over several hundreds of $nm$ \cite{Cro,Crrs,nast,DnUr,Harris,Perina,Saleh,HbGr}.  It has been shown that photon pairs generated in such structures can posses quantum temporal correlations at the timescale of fs.  Other applications of
ultra-wideband biphotons include nonclassical metrology
\cite{Giov} and large bandwidth quantum information processing
\cite{Khan}, \cite{Law}.

We consider aperiodically poled crystals as a multilayered
structure consisting of $N$ layers with variation of lengths: the
length of each layer is larger from the previous by a chirp
parameter $\zeta$. So the length of $n$-th layer is given by the
following expression $l_{n}=l_{0}+(n-1)\zeta$, where $l_{0}$ is
the length of the first layer. We consider the quadratic
nonlinearity $\chi _{n}$ having a constant
$\chi_{n}=(-1)^{(n-1)}\chi_{0}$ value within each  $n$-th layer .
This structure can be divided into $N/2$ domains such that each of them will be consist of two layers  with reversed crystal axes. So the poling period
$\Lambda$ which is the length of the domain races by $2\zeta l$
for any next domain, which means it is dependent on the
coordinate. Thus, it is easy to realize that  the relation between
chirp parameters $\zeta$ and $\alpha$ can be written as follows
$\zeta=\frac{\alpha l_{0}^{3}}{\pi}$.

 According to the formulas
(\ref{KlyshkoEq})  we obtain two-photon spectral amplitude $F(\Delta k)$
for the case of aperiodically poled crystal in the following form

\begin{widetext}
\begin{eqnarray}
F(\Delta k)=\frac{\chi_{0}}{\Delta k}\exp\left(-i\frac{\Delta
kl_{0}}{2}\right)\sum_{m=1}^{M}(-1)^{m}\exp\left(-i\Delta
k\left(ml_{0}+\frac{(m-1)^{2}\zeta}{2}\right)\right)\sin\left(\frac{\Delta
k(l_{0}+(m-1)\zeta)}{2}\right),\label{SpectrAmp2}
\end{eqnarray}
\end{widetext}
through the discrete Gauss sum,  where $\Delta k$ is the phase
mismatch function.
 In this case the probability of biphoton generation as a function of the frequency  $\Omega$  is proportional to
\begin{widetext}

\begin{eqnarray}
|F(\Omega)|^{2}=\frac{\chi_{0}^{2}}{\left(D\Omega+\Delta
k_{0}\right)^{2}}
\Bigg[\sum_{m=1}^{N}\sin^{2}\left((D\Omega+\Delta
k_{0})\frac{(l_{0}+\zeta (m-1))}{2}\right)
+2\sum_{m=1}^{N-1}\sum_{p=1}^{N-m}(-1)^{p}\times~~~~~~~~~~~~~~~~~~~~~~\\\nonumber
\times\sin\left((D\Omega+\Delta k_{0})\frac{(l_{0}+\zeta
(m+p-1))}{2}\right) \sin\left((D\Omega+\Delta
k_{0})\frac{(l_{0}+\zeta (m-1))}{2}
\right)\cos\left(p(D\Omega+\Delta k_{0})\left(l_{0}+\frac{\zeta
(2m+p-2)}{2}\right)\right)\Bigg].\label{CorFunc2}
\end{eqnarray}
Here  $\Delta
k_{0}=k_{i}(\frac{\omega_{0}}{2})+k_{s}(\frac{\omega_{0}}{2})-k_{0}$
 is the phase mismatch vector at the central idler and signal
frequencies $\omega_{i}=\omega_{s}=\frac{\omega_{0}}{2}$, where
$\omega_{0}$ is the laser pump frequency. In this formula we
have considered also $\Delta k=\Delta k_{0}+D\Omega$.
\end{widetext}

It is easy to realize that the biphoton spectra for two  systems
under consideration are essentially different in form. The main
difference is that   biphoton spectra for aperiodically poled
structures do not consist from  separated spectral peaks in the
case of small layers in opposite to the case of photonic crystal.
If the  chirp parameter $\zeta$ equals to zero the spectrum (11)
is reduced to the  spectral function of biphotons for pure
periodically poled configuration that includes segments of length
$l_{0}$ with positive $\chi_{0}$ and negative $-\chi_{0}$
susceptibilities that alternate one to the other

\begin{equation}
|F(\Omega)|^{2}=l_{0}^{2}\chi_{0}^{2}\frac{sin^{2}\left(\frac{N(D\Omega+\Delta
k_{0}-\frac{\pi}{l_{0}})l_{0}}{2}\right)}{sin^{2}\left(\frac{(D\Omega+\Delta
k_{0}-\frac{\pi}{l_{0}})l_{0}}{2}\right)}sinc^{2}\left(\frac{(D\Omega+\Delta
k_{0})l_{0}}{2}\right)\label{PerPol}
\end{equation}
In the approximation $N>>1$ this probability reads as
$F|\Omega|^{2}\sim
l_{0}^{2}N^{2}sinc^{2}\left(\frac{l_{0}(D\Omega+\Delta
k_{0})}{2}\right)$.  We illustrate the peculiarities of the biphoton
spectra in dependence on $\Omega$ for aperiodically poled
structure as well as for periodically poled structure on Figs. 2
for $N=50$ and $\Delta k_{0}=0$. As we see, in the limit $\zeta=0$
and $N>>1$ there occur only one narrow pick for positive $\Omega$
centered at $\frac{\pi}{l_{0}}$.  Increasing of the chirp parameter
broads the frequency spectrum.

\begin{figure}

\includegraphics[width=9cm]{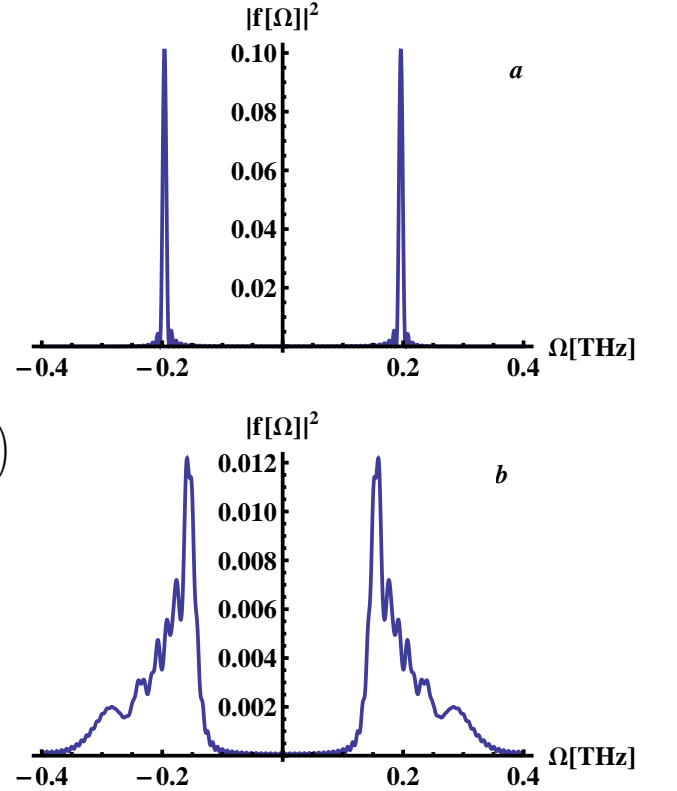}

\caption{Biphoton spectra in dependence on $\Omega$ frequency for
non-chirped QPM grating ($\zeta=0$, $l=160\mu m$ (a))  and chirped QPM grating ($\zeta=2.82$,  $l_{0}=88.09\mu m$ (b)). The parameters are: $L=8000 \mu
m$, $\lambda_{0}=0.458\mu m$, $B=0.3$, $N=50$. } \label{ch0}
\end{figure}

\begin{figure}

\includegraphics[width=9cm]{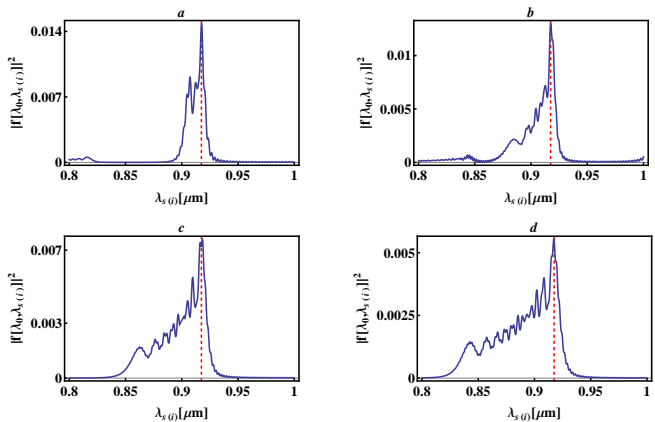}

\caption{Biphoton spectral densities in dependence on
wavelength for aperiodically poled crystals with the following
parameters: $N=50$, $l_{0}=109.5\mu m$, $\zeta=1 \mu m$, $\Delta
k_{0}=\frac{\pi}{l_{37}}$ (a); $N=50$, $l_{0}=88.09\mu m$,
$\zeta=2.82 \mu m$, $\Delta k_{0}=\frac{\pi}{l_{40}}$ (b);  $N=100$,
$l_{0}=52.225 \mu m$, $\zeta=0.55 \mu m$, $\Delta
k_{0}=\frac{\pi}{l_{85}}$ (c); $N=160$, $l_{0}=35.51 \mu m$,
$\zeta=0.18 \mu m$, $\Delta k_{0}=\frac{\pi}{l_{140}}$ (d). Here
$\l_{n}=l_{0}+n\zeta$) and $B=0.3$, $\lambda_{0}=0.458\mu m$
 and red dashed lines correspond to the wavelength
$2\lambda_{0}$}\label{ch1}
\end{figure}

Typical biphoton spectra for aperiodic, chirped  poling are shown
in Figs. \ref{ch1}   in dependence on the signal-field
(idler-field) wavelength ($\lambda_{s}\gtrsim 2\lambda_{0}$) for
various layer numbers and chirping parameters. For comparison of
these spectra with analogous results  obtained for the case of
photonic crystals as well as with the experimental results (see,
for example, \cite{kitaeva}) we use the parameters that are
suitable for the LiTaO$_{3}$  aperiodically poled crystal, with
$B=0.3$,  the whole length $L=0.8 cm$ and for the laser pump
frequency $\lambda_{0}=0.458\mu m$.

 Biphoton spectral density  for an aperiodically poled
crystal with layers number N=50, chirp parameter $\zeta=1$ and
$l_{0}=109.5$,
 which corresponds to a smaller
spatial chirp parameter $\alpha=240cm^{-2}$ is shown in the Fig.
\ref{ch1}(a). The  analogous result  for the case of crystal with
the same number of layers, but with chirp parameter $\zeta=2.82$
and $l_{0}=88.09$, which corresponds to the spatial chirp
parameter $\alpha=1200cm^{-2}$ is shown in  Fig. \ref{ch1}(b).
 From comparison of
these two results we can see that increasing of chirp parameter on
two times lead to wide  broadening of the signal- and
idler-field spectra. Moreover,  for each number of layers there is
an optimal value of chirp parameter for which the spectrum is the
broadest. In Fig. \ref{ch1}(b,c,d), we show the results of
forming the  biphoton spectra as number of layer increases
provided that the chirping parameter is fixed. For this goal we
consider the number of layers: $N=50$, $N=100$ and $N=160$ and the
chirping parameters are chosen so that again $\alpha=1200cm^{-2}$. As
we can see from these pictures the spectra  in these three cases
have approximately the same form  and they qualitatively match
with the result for the biphoton spectra broadening presented in
\cite{SenHar} but are different in details.  The wider broadening occurs for the higher number
of layers: for the number of layers $N=160$ the spectral width
achieves 0.2$\mu m$. Thus, we conclude  that for the case of
aperiodic chirping the width of spectrum is strongly dependent on
the number of layers  despite the refractive index frequency chirp
considered in the previous section. We also observe  that in the
case of aperiodically poled crystals the spectral lines
corresponding definite segments are not resolved as it is
demonstrated  in the case of photonic crystals.

\section{Conclusion}

In conclusion, we have investigated production of biphoton wave packet in two
chirped QPM
structures  presented as the ensemble of  second-order nonlinear layers.
In this approach the total amplitude of two-photon generation has been  calculated
as the superposition of partial amplitudes corresponding to each layer with local
 spatial chirp frequency. In this way, the spectrum of spontaneously generated
 chirped photons has been calculated through Gauss sums that involve realistic
 parameters  of three-wave interaction in each layers. The detailed
calculations have been done for two QPM structures: (a) chirped nonlinear
one-dimensional multi-layered photonic crystals; (b) aperiodic poled
one-dimensional multi-layered crystals.
 The advantage of the presented approach is that we have found out
 clear and easy correlation between number of layers the crystal consists
 of and the formation of broadband interference shape of the biphoton
 spectral rate.  The physical mechanism of chirping for both structures
 have been also explained  for physically realised parameters.  This approach
 allows us to consider the case of small
number of layers. Particularly, it has been demonstrated that  biphoton
spectra for chirped photonic crystals with  small number of layers  (N=5,10)
consist from well-resolved spectral lines. It has been demonstrated two mechanisms
for forming the broadband spectra of signal (idler) waves in SPDC for both
configurations as number of layers increases and for various chirping parameters.

\end{document}